\documentclass{emulateapj}

\newcommand{\simlt}{\,\hbox{\lower0.6ex\hbox{$\sim$}\llap{\raise0.2ex\hbox{$<$}}}\,}
\newcommand{\simgt}{\,\hbox{\lower0.6ex\hbox{$\sim$}\llap{\raise0.2ex\hbox{$>$}}}\,}
\newcommand{\fras}{\mbox{\ensuremath{.\mkern-4mu^{\rm s}}}}
\newcommand{\fdeg}{\mbox{\ensuremath{.\mkern-4mu^{\circ}}}}

\newcommand{\fsec}{\mbox{\ensuremath{.\mkern-4mu^{''}}}}

\slugcomment{}

\shorttitle{GRB~060121: Short Burst at High $z$}
\shortauthors{de Ugarte Postigo et al.}

\begin{document}

\title{GRB~060121: Implications of a Short/Intermediate Duration $\gamma$-Ray Burst
at High Redshift}

\author{A.~de~Ugarte Postigo\altaffilmark{1}, 
A.J.~Castro-Tirado\altaffilmark{1}, 
S.~Guziy\altaffilmark{1,2}, 
J.~Gorosabel\altaffilmark{1}, 
G.~J\'ohannesson\altaffilmark{3}, 
M.A.~Aloy\altaffilmark{4,5}, 
S.~McBreen\altaffilmark{6}, 
D.Q.~Lamb\altaffilmark{7},
N.~Benitez\altaffilmark{1},
M.~Jel\' inek\altaffilmark{1}, 
S.B.~Pandey\altaffilmark{1}, 
D.~Coe\altaffilmark{1}, 
M. D. P\'erez-Ram\'irez\altaffilmark{8}
F.J.~Aceituno\altaffilmark{1}, 
M.~Alises\altaffilmark{9}, 
J.A.~Acosta-Pulido\altaffilmark{10}, 
G.~G\'omez\altaffilmark{10}, 
R.~L\'opez\altaffilmark{11}, 
T.Q.~Donaghy\altaffilmark{7}, 
Y.E.~Nakagawa\altaffilmark{12}, 
T.~Sakamoto\altaffilmark{13}, 
G.R.~Ricker\altaffilmark{14}, 
F.R.~Hearty\altaffilmark{15}, 
M.~Bayliss\altaffilmark{7}, 
G.~Gyuk\altaffilmark{7} and
D.~G.~York \altaffilmark{7}}

\altaffiltext{1}{Instituto de Astrof\' isica de Andaluc\' ia (IAA-CSIC), 
Camino Bajo de Hu\'etor, 50, E-18008 Granada, Spain.}
\altaffiltext{2}{Nikolaev State University, Nikolska 24, 54030 Nikolaev, Ukraine.}
\altaffiltext{3}{Science Institute, University of Iceland, Dunhaga 3, IS-107 Reykjavík, Iceland.}
\altaffiltext{4}{Max-Planck-Institut f\"ur Astrophysik, D-85741, Garching bei M\"unchen, Germany.}
\altaffiltext{5}{Departamento de Astronom\' ia y Astrof\' isica, Universidad de Valencia, E-46100 Burjassot, Spain.}
\altaffiltext{6}{Max-Planck-Institut f\"{u}r extraterrestrische Physik, D-85748 Garching bei M\"unchen, Germany.}
\altaffiltext{7}{Department of Astronomy and Astrophysics, University of Chicago, Chicago, Illinois 60637, U.S.A.}
\altaffiltext{8}{Dpto. de F\' isica (EPS), Universidad de Ja\'en, E-23071 Ja\'en, Spain.}
\altaffiltext{9}{Calar Alto Observatory, Apartado de Correos 511, E-04080 Almer\' ia, Spain.}
\altaffiltext{10}{Instituto de Astrof\' isica de Canarias, V\' ia L\'actea s/n, E-38200 La Laguna - Tenerife, Spain.}
\altaffiltext{11}{Departament d'Astronomia i Meteorologia, Universitat de Barcelona, Av. Diagonal 647,E-08028 Barcelona, Spain.}
\altaffiltext{12}{Department of Physics and Mathematics, Aoyama Gakuin University, Fuchinobe 5-10-1, Sagamihara, Kanagawa 229-8558, Japan.}
\altaffiltext{13}{NASA Goddard Space Flight Center, Greenbelt, Maryland 20771, U.S.A.}
\altaffiltext{14}{MIT Kavli Institute, Massachusetts Institute of Technology, 70 Vassar Street, Cambridge, Massachusetts 02139, U.S.A.}
\altaffiltext{15}{Center for Astrophysics and Space Astronomy, University of Colorado, Boulder, CO 80303 U.S.A.}

\begin{abstract}
Since the discovery of the first short-hard $\gamma$-ray burst afterglows 
in 2005, the handful of observed events have been found to be embedded 
in nearby (z $<$ 1), bright underlying galaxies.  We present multiwavelength observations of the 
short-duration burst GRB~060121, which is the first observed to clearly outshine 
its host galaxy (by a factor $>$ 10$^{2}$).  A photometric redshift for this event 
places the progenitor at a most probable redshift of z  = 4.6, with a less 
probable scenario of z = 1.7. In either case, GRB 060121 could be the farthermost 
short-duration GRB detected to date and implies an isotropic-equivalent energy 
release in gamma-rays comparable to that seen in long-duration bursts.
We discuss the implications of the released energy on the nature of the progenitor. 
These results suggest that GRB 060121 may belong to a family of energetic 
short-duration events, lying at z $>$ 1 and whose optical afterglows would 
outshine their host galaxies, unlike the first short-duration GRBs observed in 
2005. The possibility of GRB~060121 being an intermediate duration burst is also
discussed. 
\end{abstract}





\section{Introduction}

Since 1993 $\gamma$-ray bursts (GRBs) GRBs have been classified into two subgroups 
according to the observed duration and hardness-ratio derived from their gamma-ray 
spectra: short-hard (SGRB) and long-soft (LGRB) events \citep{kou93}.

The first X-ray detections of SGRB afterglows by {\it Swift} \citep{geh05} and {\it HETE-2} \citep{hjo05,vil05}
in combination with optical and radio detections have suggested that they 
release less energy than the LGRBs \citep{fox05} (E$_{\gamma}$(SGRB) $\sim$ 10$^{49}$ erg vs. 
E$_{\gamma}$(LGRB) $\sim$ 10$^{51}$ erg) and that they may originate from the coalescence of 
neutron star (NS) or NS-black hole (BH) binaries at cosmological distances. Until 
GRB 060121, the detected optical afterglows of SGRBs have been comparable in 
brightness to their host galaxies or dimmer, unlike the case of long-soft 
bursts that may outshine their hosts by a factor of up to $\sim$ 10$^6$ \citep{van00}. The 
distance scale is also considered to be different, having all short bursts with 
definite redshifts been $z_{{\rm SGRB}}<0.6$\footnote[\dag]{We point out that there is still
an ongoing debate about the redshift of GRB~050813, which could lie at a
redshift of $z \sim 1.8$ \citep{ber06}, although no optical or radio counterpart was found
to support this scenario.} while long bursts show a broader 
range, mainly between 0.5 and 5, with $<z_{{\rm LGRB}}>  \sim 1.8$. 

\section{Observations and Data Reduction}


GRB~060121 was detected at 22:24:54.50 UT on 21 January 2006 by 
FREGATE, WXM and SXC instruments aboard the {\it HETE-2} mission \citep{ari06}. The position of 
the gamma-ray event was distributed using the GRB Coordinates Network (GCN) 
13 s later. The spectral peak energy $Ep = 120 \pm 7$ keV \citep{boe06}, together with a 
duration of $1.97 \pm 0.06$ s initially classified it in the 
SGRB group of events (Fig. 1). It was also observed by Konus/{\it WIND} \citep{gol06a}
and followed up by {\it Swift}/XRT \citep{man06}, whose observations substantially improved the 
initial 28$'$ {\it HETE-2} error box radius and helped to identify the optical counterpart \citep{mal06}.

\begin{figure}
\centering
\includegraphics[angle=270,scale=0.25]{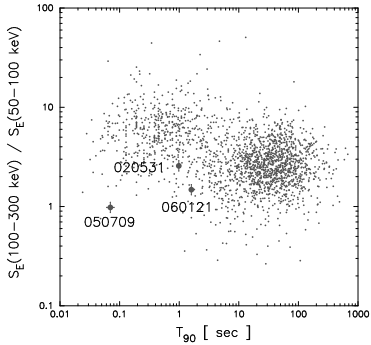}
\caption{Burst duration vs. spectral hardness diagram. Large black circles are
the locations of the three currently established {\it HETE-2} short GRBs
superimposed on the distribution of 1973 BATSE short and long GRBs (small grey
dots). \label{fig1}}
\end{figure}

On receipt of the initial alert by {\it HETE-2}, a mosaic of images was triggered at 
the 1.5 m telescope of Observatorio de Sierra Nevada (OSN) in order to map the 
entire error box. The first detection of the afterglow was obtained only 22 
minutes after the $\gamma$-ray event. Complementary observations were requested with 
the 2.2 m telescope of the German-Spanish Calar Alto Observatory (CAHA), the 
4.2 m William Herschel telescope (WHT) at Roque de los Muchachos Observatory 
and the 3.5 m Astrophysical Research Consortium (ARC) telescope at Apache 
Point Observatory. Equatorial coordinates of the optical near-infrared (nIR) afterglow yielded: 
R.A.(J2000)=$09^h09^m52\fras02$, Dec.(J2000)=$+45^\circ39'45\fsec9$
($0\fsec5$ uncertainty at 1$\sigma$ level).

Optical images have been photometrically calibrated against
Sloan Digital Sky Survey \citep{ade06} applying the corresponding transformations
\citep{jes05} for our photometric system. For nIR images we have used field 
stars from the 2 Micron All Sky Survey catalogue \citep{skr06}. A Galactic extinction
\citep{sch98} correction of E(B-V)=0.014 is applied and magnitudes have been converted to 
flux density units (Jy) for clarity \citep{fuk95,cox00}. The photometric data 
are displayed in Table 1.

\begin{table}
\begin{center}
\caption{Observations of the afterglow of GRB~060121\label{observations}}
\begin{tabular}{ccccc}
\tableline\tableline
Mean Date &  Band & Tel.    & Int. time  & Flux         \\
(Jan 2006)&       &         &   (s)      & ($\mu$ Jy)    \\
\tableline
22.2443        & K     & 4.2mWHT & 750           & 17.1$\pm$1.4        \\
23.2402        & K     & 4.2mWHT & 1000          &  6.3$\pm$1.6        \\
23.3308        & K     & 3.5mARC & 3600         & 7.48$\pm$0.65        \\
27.2588        & K     & 3.5mARC & 3600         & $<$2.13               \\
\hline
21.9495        & I     & 1.5mOSN & 120           & 19.3$\pm$4.4        \\
21.9633        & I     & 1.5mOSN & 120           & 10.0$\pm$2.8        \\
22.0458        & I     & 1.5mOSN & 300           & 10.8$\pm$2.5        \\
22.1162        & I     & 1.5mOSN & 6$\times$300         & 4.55$\pm$0.83        \\
22.2539        & I     & 1.5mOSN & 5$\times$300         & $<$2.5               \\
\hline
22.0493        & R     & 2.2mCAHA & 600                  &  3.70$\pm$0.97        \\
22.0963        & R     & 2.2mCAHA & 2$\times$600         &  1.14$\pm$0.33        \\
22.1590        & R     & 2.2mCAHA & 2$\times$600         &  1.23$\pm$0.49        \\
22.2385        & R     & 2.2mCAHA & 3$\times$600         &  1.35$\pm$0.37        \\
23.1828        & R     & 1.5mOSN  & 12$\times$900          &  $<$0.8        \\
24.0804        & R     & 2.2mCAHA & 6$\times$900         &  0.55$\pm$0.14        \\
\hline
22.0885        & V     & 2.2mCAHA & 2$\times$600    &  $<$1.1        \\
22.1952        & V     & 2.2mCAHA & 5$\times$600         & $<$0.7               \\
\hline
22.0336        & B     & 2.2mCAHA & 600           & $<$1.7               \\
22.1560        & B     & 2.2mCAHA & 7$\times$600         & $<$0.9               \\
\hline
22.0258        & U     & 2.2mCAHA & 600           & $<$3.3               \\
22.1200        & U     & 2.2mCAHA & 5$\times$600         & $<$2.1               \\
\tableline
\end{tabular}
\end{center}
\end{table}

The X-ray light curve and spectrum were obtained from the {\it Swift}/XRT data. The 
spectrum was fitted with a power-law plus a fixed Galactic hydrogen column 
density ($N_{{\rm H (Gal)}} = 1.7\times10^{20}$ cm$^{-2}$) and an intrinsic column density $N_{{\rm H (int)}}$ 
at a varying redshift in the range $0.1 \leq z \leq 6.0$ . A spectral index ($F \sim \nu^{-\beta}$) 
of $\beta = 1.10_{-0.16}^{+0.17}$ ($\chi^2/d.o.f. = 36/29$) was derived from the fit with an 
intrinsic column density that, depending on the selected redshift scenario, 
ranges from $N_{{\rm H (z = 1.5)}} = 0.39_{-0.20}^{+0.17}\times10^{22}$ cm$^{-2}$ to 
$N_{{\rm H (z = 4.6)}} = 2.9_{-1.5}^{+1.3}\times10^{22} $cm$^{-2}$.




\section{Results}
The detections in I, R and K bands and the upper limits imposed for the U, B 
and V bands, allowed us to construct a spectral flux distribution (SFD) at a 
mean epoch of 2.5 hours after the burst (Fig. 2). The nIR K band point is 
extrapolated from a near epoch using as
reference a quasi-simultaneous R passband detection.
The data were fit with a power law spectrum, a superposed intrinsic 
extinction \citep{pei92} and a Lyman-$\alpha$ blanketing 
model \citep{mad95} at a varying redshift. The slope of the powerlaw was chosen
to be $\beta_{{\rm opt}}=0.60\pm0.09$, as derived from the X-ray spectra, assuming 
$\nu_{{\rm opt}}<\nu_{\rm c}<\nu_{\rm X}$ at a pre-break epoch, following the standard fireball 
model prescription \citep{sar99}. This is confirmed by the modelling described bellow. 
We obtained two probability peaks in our 
redshift study. The main one (with a 63\% likelihood) places the burst at 
$z = 4.6 \pm 0.5$ with an intrinsic extinction of $A_{\rm V} = 0.3 \pm 0.2$ magnitudes. A 
secondary peak (with a 35\% likelihood) would imply that the afterglow lies at 
a $z = 1.7 \pm 0.4$ and $A_{\rm V} = 1.4 \pm 0.4$ magnitudes. In either case, GRB 060121 is 
the farthermost short duration GRB detected to date. A redshift of $z < 0.5$ 
has a likelihood $\leq 0.5\%$ and implies extinctions $A_{\rm V} > 2.0$. Adopting a flat 
cosmology with $\Omega_\Lambda = 0.73$, $\Omega_{\rm M} = 0.27$ and $H_{\rm 0} = 71$ 
km s$^{-1}$ Mpc$^{-1}$ and considering a $\gamma$-ray fluence of 
$4.71_{-0.31}^{+0.44}\times10^{-6}$ erg cm$^{-2}$, we obtain an isotropic-
equivalent energy release in $\gamma$-rays of $E_{\gamma{\rm ,iso}} = 2.18_{-1.72}^{+0.21}\times10^{53}$ erg 
($E_{\gamma{\rm ,iso}} = 4.30_{-2.63}^{+0.40}\times10^{52}$ erg) for a redshift of $z = 4.6$ ($1.7$), 
2(1) orders of magnitude higher than other $E_{\gamma{\rm ,iso}}$ values determined for 
previous SGRBs (Table 2). This is comparable to the values measured 
for LGRBs \citep{fra01}.

\begin{figure}
\plotone{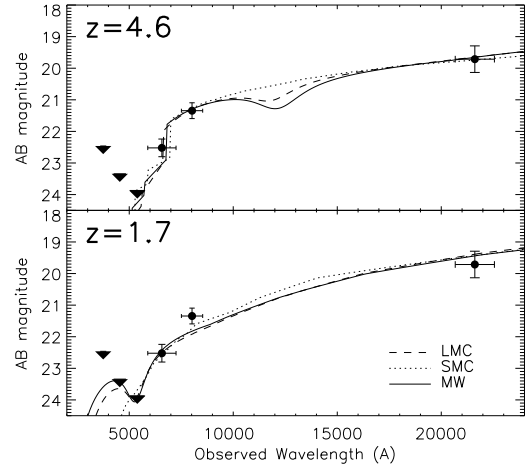}
\caption{Spectral flux distribution. A fit of the data points with an extincted power law and a
Lyman-$\alpha$ break returns two peaks of probability. The main one,
covering a probability of 63\% is centred at $z = 4.6$ with rest frame
extinction of $A_{\rm V} = 0.3 \pm 0.2$ (top) and the secondary with an 35\% is centred
at $z = 1.7$ with $A_{\rm V} = 1.1 \pm 0.2$ (bottom).\label{fig2}}
\end{figure}

\begin{table*}
\begin{center}
\caption{Physical properties of the short $\gamma$-ray bursts detected to date\label{table1}}
\begin{tabular}{ccccccc}
\tableline\tableline
GRB & Redshift $z$ & T$_{{\rm 90}}$ (s) & Fluence (erg cm$^{-2}$) & E$_{\gamma{\rm ,iso}}$
(erg) & L$_{{\rm X}}$ (erg s$^{-1}$) & F$_{{\rm A/G}}$ \\
\tableline
050509B & 0.225 & 0.04 & $9.5\times10^{-9}$ & $4.5\times10^{48}$ & $<7\times10^{41}$ & $<0.005$ \\
050709  & 0.160 & 0.07 & $2.9\times10^{-7}$ & $6.9\times10^{49}$ & $ 3\times10^{42}$ & $\sim1.0$ \\
050724  & 0.258 & 3.00 & $6.3\times10^{-7}$ & $4.0\times10^{50}$ & $ 8\times10^{43}$ & $\sim0.2$ \\
050813$^{\dag}$  & 0.722/1.8? & 0.60 & $1.2\times10^{-7}$ & $6.5\times10^{50}$ & $ 9\times10^{43}$ & $<0.15$ \\
051221A & 0.546 & 1.40 & $3.2\times10^{-6}$ & $2.4\times10^{51}$ & $ 6\times10^{44}$ & $\sim1.0$ \\
060121  & 1.5   & 1.97 & $4.7\times10^{-6}$ & $2.9\times10^{52}$ & $ 8\times10^{45}$ & $\sim20.0$ \\
        & 4.6   &      &               & $2.1\times10^{53}$ &  $6\times10^{46}$ & \\
060313  &$\leq1.7$& 0.70 & $1.4\times10^{-5}$ & $\leq1\times10^{53}$  &   -     & $>3.0$ \\
\tableline
\end{tabular}
\tablecomments{The table displays, in columns: Name of the burst, redshift,
duration of the gamma ray emission, measured gamma-ray fluence, isotropic-
equivalent $\gamma$-ray energy, isotropic-equivalent luminosity observed in X-rays 10
hours after the burst and the fraction of afterglow flux 12 hours after the
burst and the host galaxy flux (both in R band). The compilation is based on
this work and \cite{fox05,sch06,gol06b,sod06} and references therein. GRB~050813 values are 
calculated for z=0.722.}
\end{center}
\end{table*}

We have modelled the GRB 060121 afterglow following the prescription of \citet{joh06} 
on the basis of the standard fireball model for which 2 energy injections 
have been included at 0.035 and 0.23 days after the onset of the burst. These 
energy injections are required in order to explain the bumpy behaviour seen 
during the first hours (Fig. 3); similar features have been seen in long GRB 
light curves \citep{cas06} at $z \sim 4$. The model has been fitted for both $z = 4.6$ and 
$z = 1.7$ scenarios with slightly better results for the high-z case. 
Independent on the redshift, the multiwavelength model points to a narrow jet 
with half opening angle $\theta_{\rm 0} < 10^{\circ}$ in a low density environment  ($10^{-3}$ cm$^{-3} \leq n 
\leq 0.1$ cm$^{-3}$; the lower density limit fits the observations better for the case 
where the GRB took place at $z = 4.6$ while the upper density value accommodates 
better the data at $z = 1.7$), with efficiencies of conversion of kinetic to 
gamma-ray energy $\eta_\gamma < 0.05$.

\begin{figure}
\plotone{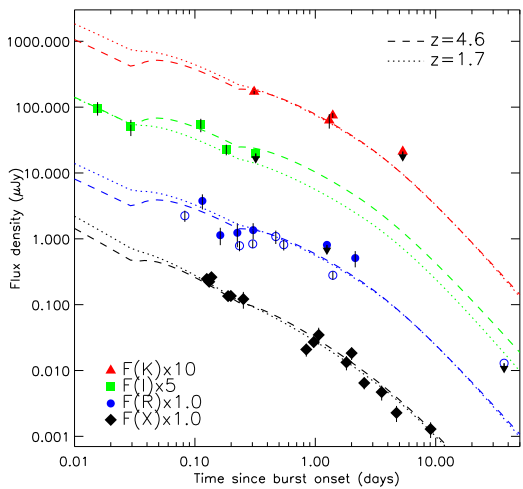}
\caption{Light curve of the afterglow in the near infrared (K), visible (R \&
I) and X-ray bands. The figure shows the result of the fit of the model in the
most probable high-z (4.6) case which has been optimized with values of
$p = 2.06$, $\theta_0 = 0\fdeg6$ and $n = 0.1$ cm$^{-3}$ and in the low-z (1.5) case, which
gives a slightly worse fit with $p = 2.05$, $\theta_0 = 2\fdeg3$ and $n = 0.04$
cm$^{-3}$. Filled symbols are data presented in this article whereas empty ones are
data from the literature.}
\end{figure}

\section{Discussion}
GRB~060121 has a $T_{90}$ duration of $1.97 \pm 0.06$ s in the 85-400 keV
energy band and a hardness ratio (HR) S$_E$(100-300 keV)/S$_E$(50-100 keV) = 1.48 $\pm$
0.18.  We note that the intrinsic duration of
GRB~060121 would be $1.6$ s with HR of $1.7$ if placed at $z \sim 0.2$ (similar to
GRB~050709), $0.7$ s and $3.3$ (if $z = 1.7$) or $0.4$ s and $4.6$ (if $z = 4.6$).
We have studied the classification of the progenitor of this GRB (as "old" or "young") 
considering eight criteria: (a) duration, (b) pulse widths, (c) spectral 
hardness, (d) spectral lag, (e) energy radiated in $\gamma$-rays, (f) existence of a 
long, soft bump following the burst, (g) location of the burst in the galaxy and 
(h) the type of host galaxy. Four (a, e, f, g) criteria provide strong evidence 
that GRB~060121 had an "old" progenitor, while four (b, c, d, h) are inconclusive. 
Thus, we can make a strong claim, although not conclusive, that GRB~060121 had an 
"old" progenitor. Further details on this analysis are given in \citet{don06}.

If we consider the progenitor to be a merger of compact objects \citep{eic89,alo05}, the 
released energy needs either a large conversion efficiency of the accreted 
mass into neutrino emission ($\simgt 0.05$), a large accretion disk mass ($\simgt 0.1 M_{\odot}$) 
or an appropriate combination of both factors \citep{oec06}. Considering a merger at a 
redshift $z > 1.5$, there is a higher consistency with the theoretical models 
where the rate of NS+NS or NS+BH mergers follows the star formation rate with 
delays of $\sim 1$ Gyr \citep{jan05} than with those of very old populations of NS+NS 
progenitor systems \citep{nak05}. Furthermore, the high extinction derived from the SFD 
fit indicates that the merger probably took place within the galaxy rather 
than in the outer halo or intergalactic medium and thus, the system received 
a small natal kick, although the density of the local event environment is 
relatively low ($n \sim 0.1$ cm$^{-3}$) as indicated by the afterglow fits discussed 
here. Alternatively, if the energy was extracted via a Blandford-Znajek 
process \citep{bla77}, either the value of the dimensionless angular momentum of the 
central BH is $a > 0.3$, the magnetic field surrounding the BH is $B \simgt 10^{16}$ G 
or the BH  has a mass larger than $3 M_{\odot}$  (or a combination of these 
parameters).

Late observations by the {\it Hubble Space Telescope} ({\it HST}) have shown no trace of 
the afterglow down to magnitude $R \sim 28$ about 37 days after the burst 
but do show an underlying galaxy \citep{lev06}. We estimate the probability of
finding such a galaxy within the $0\fsec5$ radius error box (derived from our
astrometry) as $1.3 \times 10^{-2}$ following \citet{pir02}. A chance association is unlikely 
but not completely negligible. We have reanalyzed the photometry of the galaxy 
using ColorPro \citep{coe06} obtaining $F606W_{{\rm AB}} = 27.35 \pm 0.16$ ({\it HST} wide 
band filter centred at 606 nm) and $F160W_{{\rm AB}} = 24.05 \pm 0.38$ ({\it HST} wide 
band filter centred at 1600 nm). Although the information is very limited we have used 
this photometry to study the probability distribution for the redshift of the galaxy using 
a bayesian photomeric redhift as described by \citet{ben00,ben04}. We obtain two peaks 
of probabiliy at $z=1.0$ and $z=5.2$, the latter being more probable. By multiplying
this probability by the one obtained for the afterglow we can assign a probability
of 70\% for the higher redshift case and a 28\% for the lower redshift scenario (see Fig. 4).
Applying a prior based on well determined extinctions for LGRBs \citep{kan06} the 
likelihood of a high redshift event would rise to 98\%. This is a soft constrain as it favours 
low extinctions, like it could be expected from the low density derived from our model.

\begin{figure}
\plotone{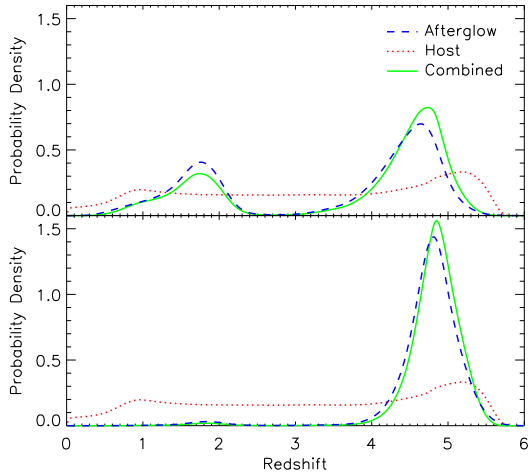}
\caption{Probability distribution vs. redshift. The top pannel shows the probability
distribution for the GRB afterglow, the host galaxy and a combination of both which 
favours the high redshift scenario. In the bottom pannel we have used an extinction-based 
prior which cancels the low redshif peak.}
\end{figure}

\section{Conclusion}
These results suggest that assuming that GRB~060121 were a SGRB, there exists 
an emerging population of short events located at high redshifts and with energies 
comparable to those of long events. In this group we may also include GRB~060313 \citep{sch06} or
GRB~000301C \citep{jen01}. This population would produce afterglows which significantly outshine 
their host galaxies, with isotropic energy releases of $E_{\gamma{\rm ,iso}} \sim 
10^{52-53}$ erg similar to the values observed in long events. Furthermore, 
following the classification by \citet{hor06} GRB 060121 could be classified in 
the “intermediate group” \citep{hor98,bal01} of events with a 68\% probability (28\% for short-burst 
and 4\% for long-burst). The relationship between this second population of short 
bursts and the “intermediate population” of GRBs will be determined or excluded by 
future observations.

\acknowledgments
We thank the generous allocation of observing time by different Time Allocation 
Committees. We acknowledge the {\it Swift} team for making 
public XRT data and S. Barthelmy for maintaining the GRB Circular Network 
that distributes the GRB alerts. We also acknowledge fruitful dicussions with 
M. Cervi\~no.

\clearpage

\end{document}